# The Adaptive Optics modes for HARMONI – From Classical to Laser Assisted Tomographic AO

B. Neichel*[a], T. Fusco[a,b], J.-F. Sauvage[a,b], C. Correia[a], K. Dohlen[a], K. El-Hadi[a], L. Blanco[a,b], N. Schwartz[c], F. Clarke[d], N. A. Thatte[d], M. Tecza[d], J. Paufique[e], J. Vernet[e], M. Le Louarn[e], P. Hammersley[e], J.L. Gach[a], S. Pascale[a], P. Vola[a], C. Petit[b], J.-M. Conan[b], A. Carlotti[f], C. Verinaud[f], H. Schnetler[c], I. Byson[c], T. Morris[g], R. Myers[g], E. Hugot[a], A. M. Gallie[c], David M. Henry[c]

[a]Aix Marseille Université, CNRS, LAM (Laboratoire d'Astrophysique de Marseille) UMR 7326, 13388, Marseille, France; [b]ONERA, 29 avenue de la Division Leclerc, 92322 Châtillon, France; [c]UK Astronomy Technology Centre, Blackford Hill, Edinburgh EH9 3HJ, United Kingdom; [d]Dept. of Astrophysics, University of Oxford, Keble Road, Oxford, OX1 3RH, United Kingdom; [e]European Southern Observatory, Karl-Schwarzschild-Strasse 2, D-85748 Garching b. Munchen, Germany; [f]Institut de Planétologie et d'Astrophysique de Grenoble, Université Grenoble-Alpes, CNRS, B.P. 53, 38041 Grenoble – France ; [g]Department of Physics, South Road, Durham, DH1 3LE, UK,

## ABSTRACT

HARMONI is a visible and NIR integral field spectrograph, providing the E-ELT's core spectroscopic capability at first light. HARMONI will work at the diffraction limit of the E-ELT, thanks to a Classical and a Laser Tomographic AO system. In this paper, we present the system choices that have been made for these SCAO and LTAO modules. In particular, we describe the strategy developed for the different Wave-Front Sensors: pyramid for SCAO, the LGSWFS concept, the NGSWFS path, and the truth sensor capabilities. We present first potential implementations. And we asses the first system performance.

**Keywords:** E-ELT, Adaptive Optics, Instrumentation, Laser

## 1. INTRODUCTION

HARMONI [1] is a visible and near-infrared integral field spectrograph, providing the E-ELT's core spectroscopic capability. It will exploit the E-ELT's scientific niche in its early years, starting at first light. To get the full sensitivity and spatial resolution gain, HARMONI will work at diffraction limited scales. This will be possible thanks to two adaptive optics systems, complementary to each other. The first one is a simple but efficient Single Conjugate AO system (good performance, low sky coverage), fully integrated in HARMONI itself. The second one is a Laser Tomographic AO system (medium performance, very good sky coverage).

HARMONI is led by a consortium of 6 partners - two from UK: Oxford University (PI institute), UK-ATC; two from Spain: IAC in Tenerife, and the CAB in Madrid; and two from France: CRAL in Lyon and LAM in Marseille. On top of that, four associate partners are also working with the HARMONI team; these are the University of Durham, ONERA (Paris), IPAG (Grenoble) and RAL Space.

HARMONI will implement four spectral scales, covering from the visible to the NIR, and from a spectral resolution of 3000 to 20000. Those spectral pixels are matched over four spatial scales, covering a Field-Of-View of ~6.5" x 9" for the biggest pixels (30x60mas projected on sky), to a narrow FoV of 0.6" x 0.9" with pixels of 4mas working at the diffraction limit of the 39m E-ELT. All the details about the instrument, and the science cases development, can be found in a partner paper presented at this conference: Thatte et al. [2].

*benoit.neichel@lam.fr;

In terms of implementation at the telescope, HARMONI will be installed on one of the Nasmyth E-ELT platform. HARMONI will be on one of two the Pre-Focal Station (PFS) side port. Fig. 1 shows a potential implementation of the instrument, where one recognizes the PFS as the grey box on the left. HARMONI is then made of several sub-components, the main one being the Integral Field Spectrograph (IFS) cryostat. This 4-meter diameter cylinder includes all the pre-optics, the Integral Field Unit (IFU) and the spectrometers. Note that this entire bloc is actually rotating in order to track the field of the view when observing. On top of this bloc, and co-rotating with it, we have the NGSWFS space, where the WFS for the SCAO, the LTAO-NGS and LTAO-Truth will be implemented. This bloc is kept at a low temperature (goal is minus 30 degree C), to minimize the thermal background. The telescope PFS provides a focal plane which is only 1 meter away for the mechanical structure of the telescope. An optical relay is therefore required to re-image the telescope focal plane onto the cryostat and for the NGS WFSs. A static, cooled modified Offner, sitting on top of the NGSWFS volume, will do this optical relay function. At the entrance of this relay, a static-warm calibration volume will be used for both AO and instrument calibration. Finally, when operating with the lasers, a dichroic is inserted at the instrument entrance, that will reflect the 589nm light into the LGSWFS. This last volume is rotating to track the telescope pupil.

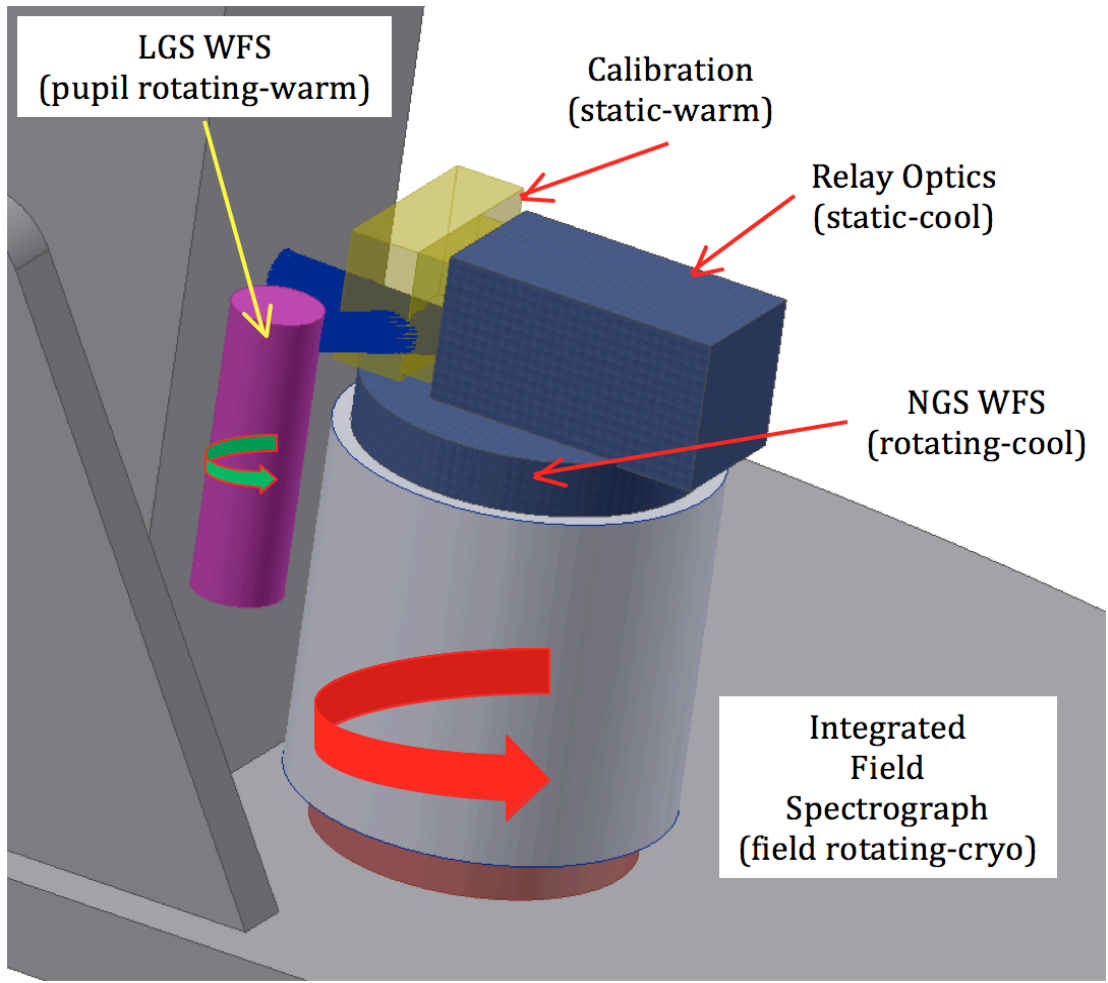

**Figure 1: Schematic view of the different sub-components of HARMONI, when installed on the side-port of the E-ELT pre-focal Station.**

In the following, we give more details on the AO modules, both the SCAO and LTAO ones.

## 2. SINGLE CONJUGATE ADAPTIVE OPTICS

The HARMONI SCAO module is designed for high performance science cases, at the price of low sky coverage and a small field of view. We are presenting in this paper the status of the preliminary design phase, 9 months after the kick-off of the whole HARMONI project.

The first subsection reminds the specifications of SCAO as known today. The next subsection is the system analysis, and lists the required functionality to achieve the performance and operation. The next subsection presents the overall error budget prediction of the SCAO system. The next subsection presents a specific point of the system, which is the telescope environment. The last subsection presents the very firsts E2E simulation results.

### 2.1 Specifications

The main performance specifications for HARMONI-SCAO module are listed below. As the final delivery of SCAO is mainly a wave-front sensor, the responsibility for Strehl ratio is jointly shared with ESO.

- For the median seeing conditions, the SCAO wave front sensor shall measure a residual wave front with an accuracy of **≤100nm RMS** for an on-axis point **source of Mv<12 (type G0)** at a temporal rate of **500Hz** and a latency of **<1.5ms**.

- Under the median seeing conditions the SCAO system shall deliver an **on-axis Strehl ratio > 0.70 at 2 μm wavelength** for AO guide stars brighter than V = 12 mag and median wind conditions.

- The SCAO wave front sensor shall be able to acquire and use a reference source **within 15-arcseconds** (goal 30-arcseconds) of the HARMONI field centre.

- The SCAO wave front sensor shall be capable of using stars or small extended objects **up to 2.5 arcsecs diameter** / FWHM (Neptune: goal up to 4 arcsecs for observations of Uranus) as the guide objects.

- Keeping the same guide star(s), the SCAO wave front sensor shall be able **to offset to up to 10 arcsec with an accuracy of 1.5 mas RMS** and be ready for science observations within 10 seconds of the telescope completing the offset.

- The SCAO wave front sensor shall be capable of accommodating a difference in guiding and observing wavelengths **up to a zenith distance of 60** (goal 70) degrees with an accuracy limited by knowledge of guide star colour and atmospheric parameters.

### 2.2 System analysis

In order to reach these ambitious specifications, the SCAO system will consists in different subsystems, shown on Figure 2.

- A SCAO dichroic picking the visible light from the NGS for WFS path. A set of 2 dichroic are decided : up to 800nm and up to 1000nm. In addition to that, a very small fraction of IR light will be propagated for differential Tip-Tilt sensing.

- A differential tip-tilt sensor, based on the IR light will ensure the measurement of NGS position and ensure a closed-loop stabilization at low framerate of the NGS on the science path.

- A WFS path with pyramid wave-front sensor, dedicated to precise atmospheric turbulence correction, as well as instrumental defect compensation and telescope perturbations correction.

- A beam steering mirror allowing to modulate the tip-tilt for pyramid at high framerate. This functionality will allow pyramid to deal with different turbulent conditions and bootstrap.

- An open-loop DM is added to modulate the static to quasi-static aberrations and allow Pyramid WFS to work around zero phase measurement.

- ADC prisms are mandatory to ensure that the image quality on the pyramid top remains of high quality.

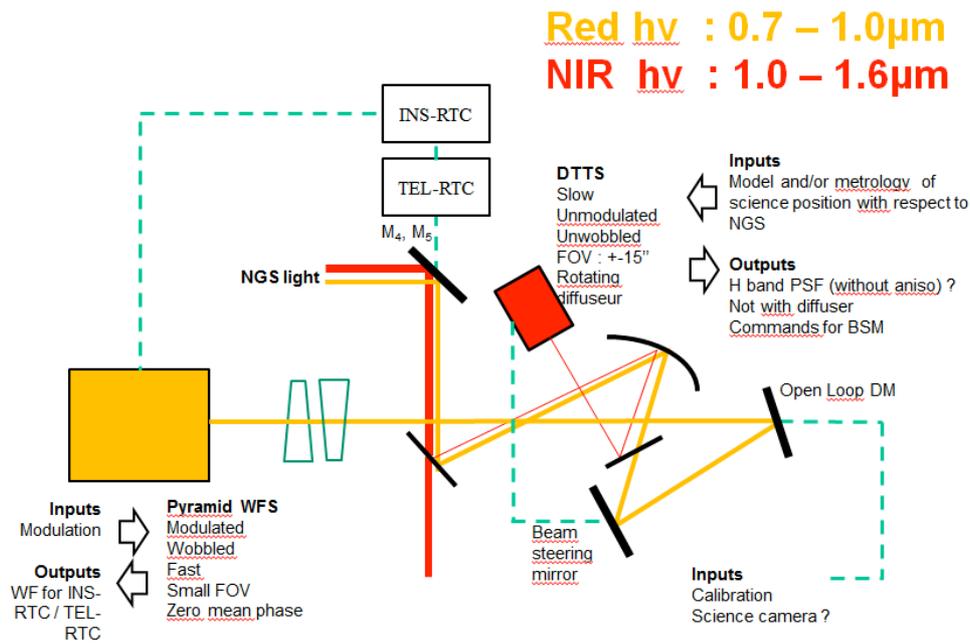

Figure 2: Preliminary sketch for HARMONI-SCAO WFS path

## 2.3 Overall error budget

The error budget has been divided into three main sections. The atmosphere (ATM) contributor includes the residual due to the atmospheric perturbation of the wave-front, evolving with time. The realistic conditions of turbulence have been obtained following ESO proposed characterization for the turbulence profile and strength. The median seeing is 0.65arcsec (at 0.5 mic), varies from 0.45 to 1.0 arcsec (optimistic – pessimistic). The instrument (INS) contributor includes the static aberrations brought by the common path (that would remain uncorrected by AO system), the non-common path uncorrected, and the quasi-static aberrations that evolved during their calibration and the observation. This error budget of 100nm in realistic case is taken from our SPHERE experience, where the final wave-front error has been estimated at 42nm RMS, with optimized optics and dedicated calibration procedures, and extrapolated to a non-XAO system.

|     | [nm RMS] | Optimistic | Realistic | Pessimistic |
|---|---|---|---|---|
| ATM | Turbulence | 120 | 152 | 207 |
| INS | Internal perf | 75 | 100 | 125 |
| TEL | Dynamic-Jitter | 47 | 65 | 121 |
|     | Dynamic-Low orders | 1 | 3 | 10 |
|     | Transient | 0 | 56 | 75 |
|     | Static | 20 | 30 | 50 |
| Total TEL | | 51 | 91 | 152 |
|     | Total (nm) | 150 | 204 | 285 |
|     | Performance (SR@K) | 83 | 71 | 51 |

Figure 3: error budget for SCAO module, including atmosphere, instrument and telescope.

## 2.4 Telescope environment

The E-ELT telescope is under design. Precise simulations of its optical quality are provided by ESO, in order to understand the environment in which the SCAO system will have to work. The telescope will be one of the critical contributor to the aberrations. First in term of amplitude (jitter due to wind shake is expected to be up to 170mas depending on wind speed / orientation and telescope orientation). But the telescope contribution will also be a complex contributor to the aberrations, by instance the transient low order optimization process will generate a quick transient tip-tilt phenomenon, every 5 minutes. The ESO data package contains tens of contributor of this kind, simulating the telescope environment. We show here in Figure 4 only the transient term (left side), for which we have tested dedicated control law in E2E mono-dimensional study, and the residual after filtering the piston shift due to wind speed on the 798 segments.

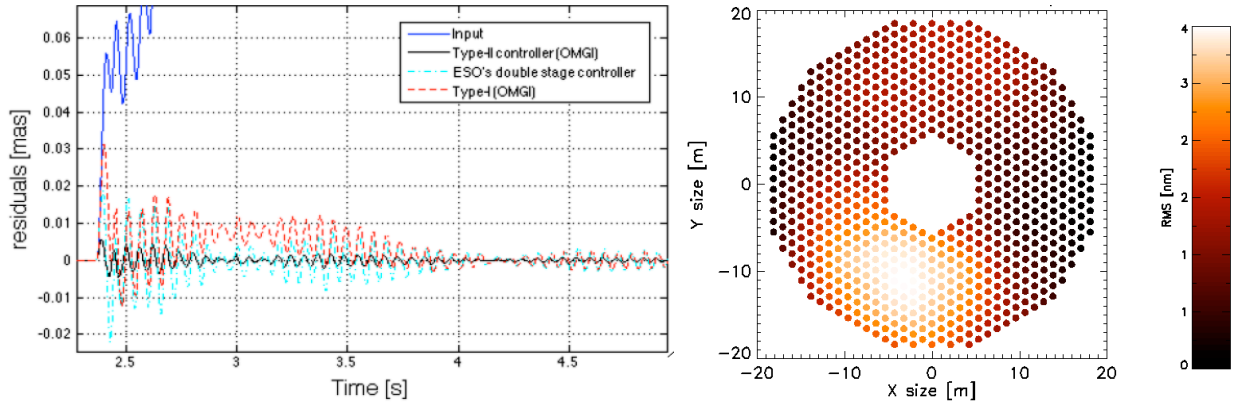

**Figure 4: [left] Transient low order optimization. The tip-tilt perturbation (solid blue) introduced by this effect is strong, and repeated every 5 minutes. Different controllers have been tested to compensate for it, and show that a mas residual can be obtain with a type-II controller (OMGI). [right] Piston induced on each M1 segment by the wind. Depending on the geometry of the problem, the contribution might not be uniform.**

## 2.5 SCAO analysis

The SCAO analysis aims at quantifying the final performance of the system, and providing the specifications for the AO components. The SCAO analysis is ongoing for HARMONI-SCAO module. The first tool used for this analysis is a fast fourier simulator enabling a quick exploration of the parameter space. The expected performance of the system with respect to magnitude is shown in Figure 5. Two kind of WFS have been compared : the SH WFS (including a 3e read-out noise detector, working at 800Hz max) and a Pyramid WFS (including a 0.3e read-out noise detector working at 1000Hz max). For the pyramid WFS, the two dichroic cases have been considered. The higher spectral cut-off gives more photon and therefore an additional 1.5 limit magnitude in mid-performance regime. Globally, the pyramid allows for a ~3magnitude gain at high performance level, or to have a better performance at a given magnitude. The baseline choice for WFS is therefore pyramid, allowing a better sensitivity.

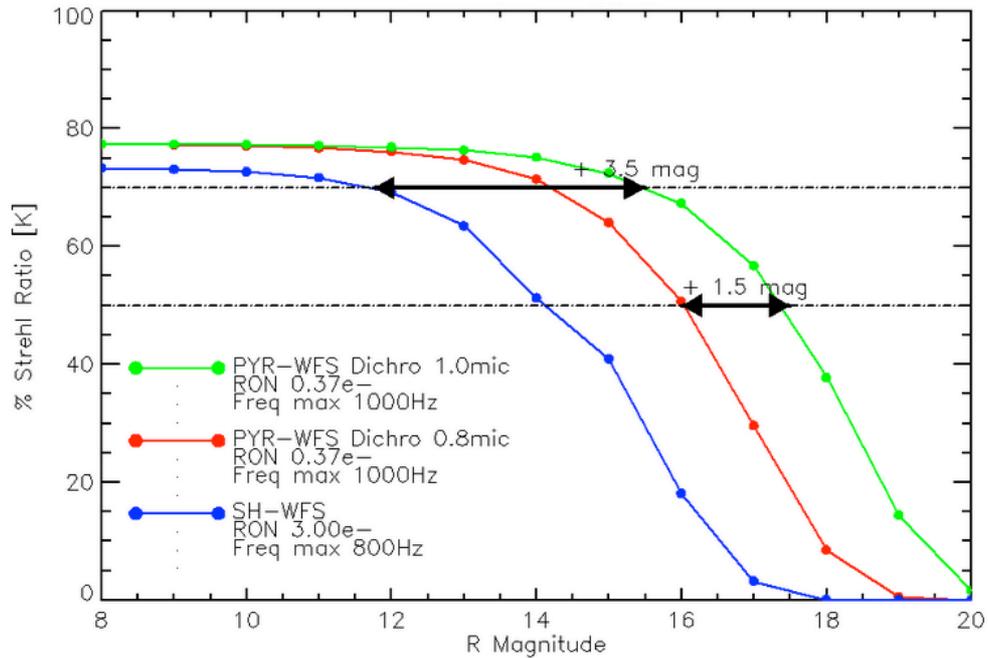

**Figure 5: expected performance of HARMONI-SCAO module, with respect to the guide star magnitude. The comparison of SH and Pyramid performance pushes forward to the baseline choice of pyramid, for is higher limit magnitude.**

The overall performance shown by fourier analysis is unable to estimate the performance of the instrument including the complex telescope environment, and most of the detailed error source of the AO and instrument itself. A second tool used for this study is therefore an end-to-end tool, matlab-based AO simulator Object Oriented Matlab Adaptive Optics (OOMAO [1]). We have developed in this tool the functionality required for simulating the HARMONI-SCAO environment (pyramid WFS, modulation, telescope environment including M4 mirror geometry and actuators, instrument internal defects).

This tool is used today and will shortly produce the overall system PSF and performance.

## 2.6 High-Contrast capability

Along with the SCAO design, high-contrast capability is currently explored in the instrument PDR. The performance goal is to provide a contrast of 1e-6 at 200mas from the optical axis. This study is led by IPAG [3], and the current potential implementation involves the use of pupil apodizers, as no intermediate focal plane will be available to insert a coronagraph. First simulations, including realistic assumption for the optics and non-common path aberrations, and the SCAO performance presented above, show that such a contrast may be achieved. Fig. 5 shows a potential mask for the apodizer, and the resulting PSF on the right. These results are still under consolidation, and more information can be found in a dedicated paper: Carlotti et al. [3].

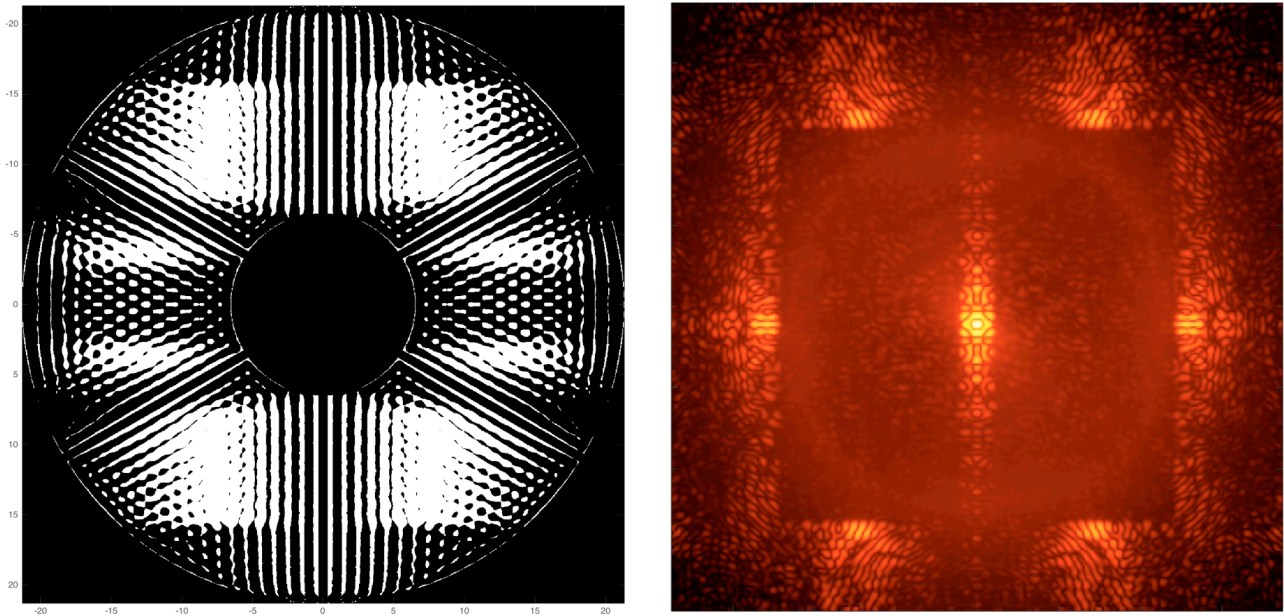

**Figure 6: left - pupil apodizer shape. Right: resulting PSF**

## 3. LASER TOMOGRAPHY ADAPTIVE OPTICS

The LTAO system of HARMONI is designed to provide the best sky coverage out of the instrument. It is a natural complement to the SCAO system, when no bright enough GS may be found. As such, it is expected that the LTAO mode would be one the most demanded for HARMONI.

The Top-Level Requirements (TLR) are currently under study, however, we have assumed based on previous requirements for ATLAS [4], that mainly two modes of operations would be relevant for the LTAO: one providing very high performance, i.e. working with the finest pixels of HARMONI, and optimizing the Strehl Ratio, however over a limited, but descent fraction of the sky; and one providing a larger sky coverage, for however some reduced sky coverage. Again, the TLR are still under definition, but a potential set of values could be close to:

- 40% EE in 40mas in H under 75%-ile seeing conditions over 60% of the sky at the South Galactic Pole
- 20% EE in 10mas in H under 75%-ile seeing conditions over 20% of the sky at the South Galactic Pole.
- 30% Strehl in H under 25%-ile seeing conditions over 10% of the sky at the South Galactic Pole.

Based on those first requirements, we have been looking at potential design implementation, with the first trade offs that are described in the following sub-sections.

### 3.1 LGS constellation

One of the first aspects we have been looking at is the LGS constellation size. HARMONI will make use of the 6 LGS provided by the telescope. Those LGSs are launched from the side of the E-ELT primary mirror. A first question that we had to face was how the performance would degrade with respect to the constellation radius, and would it be possible to separate the LGS enough so that no dichroic would be required in the instrument path, as it was the case in the previous ATLAS design. This is explained schematically in Fig. 7.

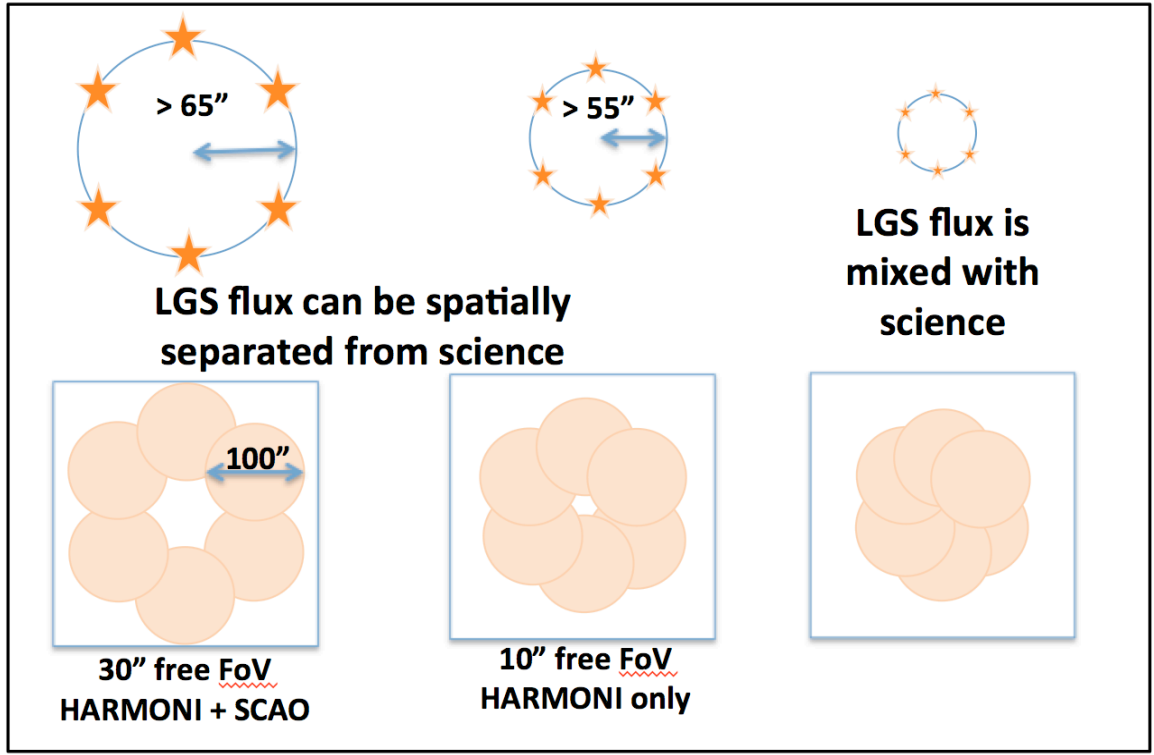

**Figure 7: LGS constellation or how to separate the LGS flux from the science photons**

To address that question, we have performed a set of first simulations in order to evaluate how the performance is scaling with the constellation radius and the zenith angle. Our Monte-Carlo simulations are performed with the Object-Oriented Matlab Adaptive Optics toolbox [5], freely available from https://github.com/cmcorreia/LAM-Public. This simulation results have also been cross-compared with results provided by Miska Le Louarn (ESO) using the Octopus tool [6]. The simulation parameters used are described in Table 1.

**Table 1: LTAO configuration and main simulation parameters**

| Parameter | Value |
| --- | --- |
| **LGS** | |
| 6 LLT | edge launch |
| Photon counts | 500ph/subaperture/frame |
| Asterism | Hexagonal – 40" radius<br>point-source projected on-sky |
| **Atmosphere** | |
| Seeing @ l=0.5 (arcsec) | JQ1  JQ2  JQ3  JQ4  Median<br>0.44  0.58  0.74  1.06  0.65 |
| L0 | 25m |
| **LO-WFS** | |

| 1xShack-Hartmann | Full-aperture, 10x10pix, Nyquist-sampled |
| --- | --- |
| wavelength | H-band (1.65 micron) |
| **HO-WFS** | |
| 6xShack-Hartmann | 74x74, 8pix/lenslet, Nyquist-sampled |
| wavelength | 589 nm |
| **DM** | |
| M4 | 4048 (all-glass), regular grid, Fried layout |
| Influence functions | Bezier, 0.4 cross coupling |
| **Controller** | |
| High-order modes | POLC + MMSE, iterative solver implementing spatio-angular tomography |
| TT | Integrator |

Results are shown in Fig. 8. For each Zenith angle, the different curves show different performance depending on the level of accuracy assumed in the tomography (see next section).

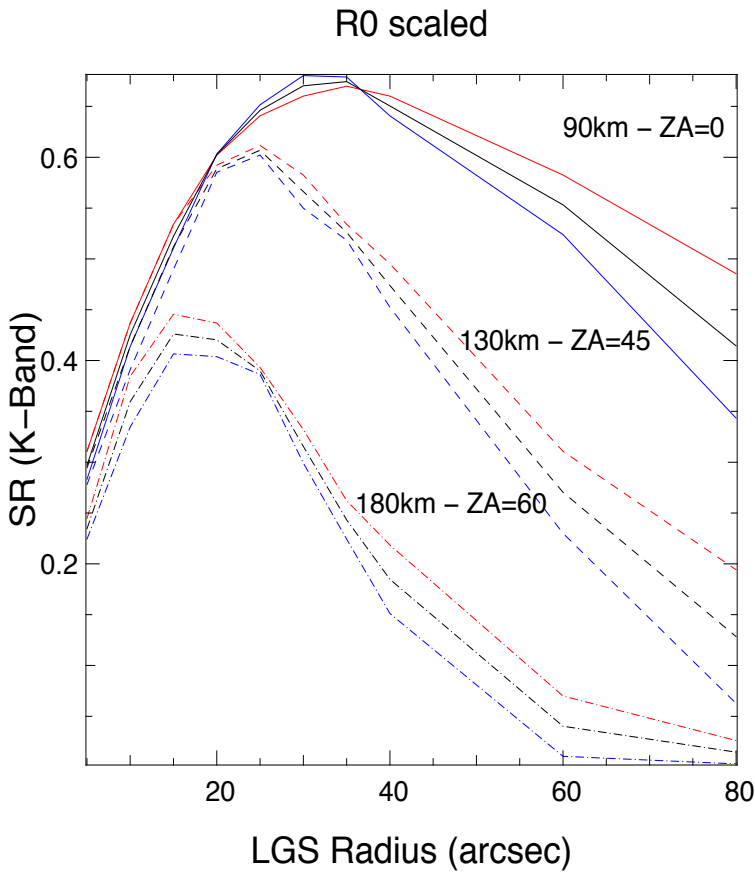

**Figure 8: LTAO performance (Strehl @ K-band) versus LGS constellation radius, for three different Zenith Angle.**

On this plot, one can see two regimes, and an optimal constellation radius, that would depend on the Zenith Angle. For constellation that would be smaller that the optimal constellation radius, the cone effect is rather large because part of the turbulence is not measured at the edge of the meta-pupil. For constellations larger than the optimal constellation, the performance degrades because of tomographic errors, unseen modes, and tomography model error. The optimal constellation angle scales as the cosine of the Zenith angle, and ranges from ~30" – 35" at Zenith, down to ~20" at 60 degree distance from Zenith. Based on these results, one of the first conclusions is that, if one wants to work around the optimal angles, there is no possibilities to physically separate the LGS flux from the science flux. Hence, a dedicated LGS dichroic is mandatory in the design of the HARMONI LTAO. The second conclusion is that, if one wants to optimize the performance, hence the LGS constellation must be adjustable depending on the where the telescope is pointing. This is the baseline that has been chosen for HARMONI. It ensures to always work at the optimal constellation angle, i.e. when the laser cones are fully synthetizing the equivalent cylinder. Note that this also made the LGSWFS design slightly simpler, as the laser will always be pointing along the telescope

"rim ray". It is also a sensible choice as the final decision is independent of the Cn2 profile used for the design. Whatever the Cn2, the best performance will be obtained for this optimal constellation. Finally, this makes the tomography much more robust to model errors, as it is described in the next section.

**3.2 Sensitivity of the tomography**

In this section, we have explored the sensitivity of the tomographic error to variations in the Cn2 profiles, and assumptions made for the reconstruction process. This study is fully described in a partner paper: Fusco et al. [7]
Basically, the idea was to explore how the tomographic error is evolving along a "typical" night. Typical Cn2 variations are obtained from models that are developed by Elena Masciadri and team [8]. As an example, Fig. 9 shows how the Cn2 would be changing along a given night. The temporal resolution on those profiles is 2 minutes, and the spatial resolution is 150 meters.

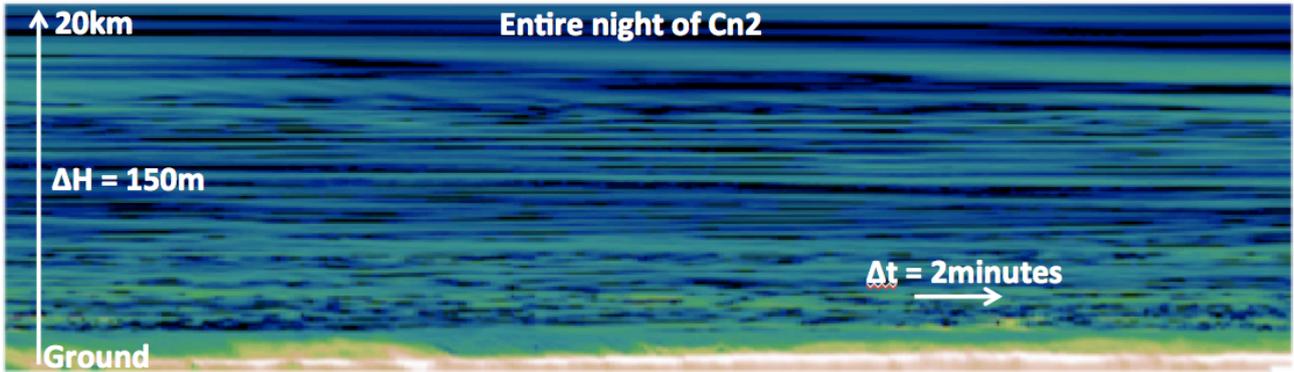

Figure 9: Cn2 (atmospheric turbulence) evolution across a full night

We then used those profiles in our tomographic simulation code, and the results are shown in Fig. 10.

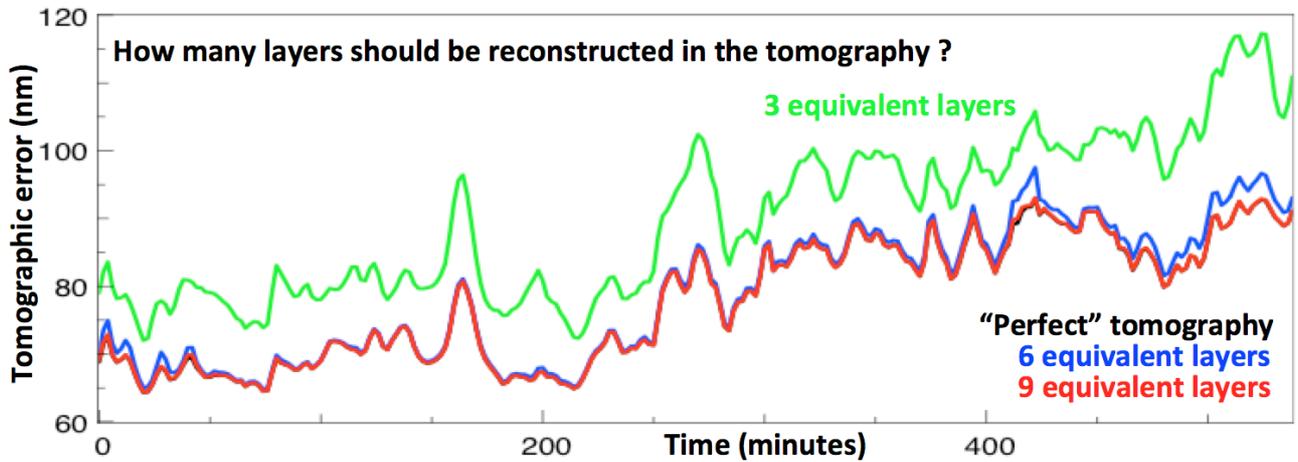

Figure 10: Evolution of the tomographic error during a typical night, for different level of tomographic reconstruction complexity.

Plots in Fig. 10 show the evolution of the tomographic error along the night, when the tomographic reconstruction is done with a perfect knowledge of the Cn2 profile at any time (black curve), and when the tomographic reconstruction is done on a given number of equivalent layers, from 3 (red), 6 (blue) and 9 (red). The variations of the black curve along the night are due to variations of the natural seeing per layer. This study shows that, at least to a first order, reconstructing on only few layers, e.g. 3, will be detrimental for the performance, while as soon as 6 or 9 layers are effectively reconstructed in the tomography, the remaining error is minimal. This tells us that a "simple" tomographic scheme may be used for HARMONI, especially because we have chosen to work around the optimal LGS constellation. Again, more details on that study, and how the performance evolves with respect to the number of LGS, or the LGS constellation size are presented in the partner paper (Fusco et al. [7]).

**3.3 Dealing with spot elongation**

The other main aspect of working with LGS on the E-ELT is the spot elongation. Indeed, as the sodium layer has a finite thickness, the LGS spots will appear to be elongated on the Shack-Hartmann subapertures. This is especially true for the E-ELT that is considering a side launch configuration.

The mean Sodium layer altitude is 90.8km above sea level [9]. The altitude of Armazones being 3046m, the mean Sodium layer altitude is 87.8km above Armazones when observing at zenith. The minimum altitude reached by the Sodium is 89km above sea level, i.e. 86km above Armazones.

The Sodium layer thickness is defined in terms of occurrences, encompassing a given number of photons [9]. If we consider 95% of the occurrences, for 95% of enclosed photon count, the Sodium layer width is 22km.

Taking into account both the Sodium mean height, and the Sodium layer width, one can derive the maximal spot elongation to be expected. This is given by the following formulae:

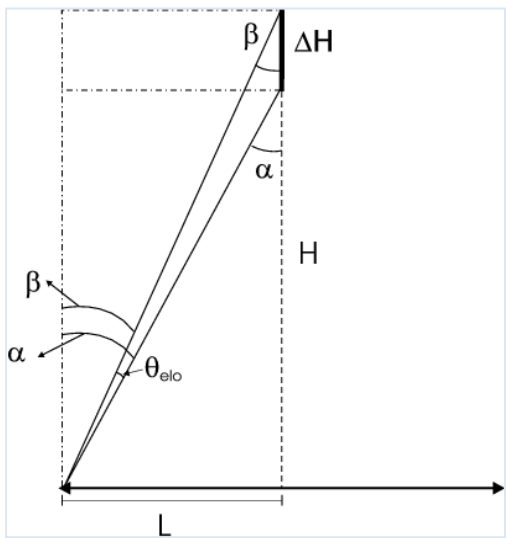

$$\gamma = \frac{L\Delta H}{H^2 + H\Delta H},$$

For a side launch (laser launch located at 22m), the maximal elongation expected at zenith is:

L = 22*2 = 44m

H = (86 – 22/2) = 75km

DH = 22km

Elongation = 27.4arcsec

And the elongation computed for the mean altitude (87.8km), gives an elongation of 26.3arcsec

Dealing with such a spot elongation requires using a large number of pixels per shack-hartmann subapertures. Indeed, if we consider that the small axis has a size of ~1", and if we want to properly sample non-elongated spots, a pixel size of 1" maximum has to be considered. On the other hand, if we want to avoid truncation effects, and get all the elongated spots, we need a FoV per subapertures of 25" at least. This leads to 25x25 pixels per sub-aperture. The current baseline being to work with 80x80 subapertures per Shack-Hartmann, it means that we need 2000 x 2000 pixels on the LGSWFS detectors. Obviously, those detectors must be read at high frame rate (> 500 Hz) and with the minimum amount of Read-out-Noise (< 3 e-). Such detectors are not available as of today, and ESO will most likely provide detectors with ~ 800 x 800 pixels. As a consequence, and if we want to keep the sampling to ~1", it means that we will have to deal with strong spot truncation.

There are several options that have been proposed to deal with spot truncation. Among others, we can cite potential solutions that has been presented during the conference:

- Retrieve the LGS induced aberrations, working with a truth sensor. Indeed, truncation will introduce a bias in the measurement, bias that will be different for all the sub-aperture, and that will propagate into the tomography to result in some given (hopefully) low order aberrations in the science path. If one can have access to a measurement of those low order aberrations with a Natural Guide Star, it may be possible to update the reference slopes of the LGSWFS in order to remove them from the tomographic reconstruction. This works as a "truth-sensor" that would always give the reference measurement. All the uncertainties around such a solution is on the spatial and temporal sampling required to keep the LGS aberrations at a low enough level [10].

- Another approach is to use advanced centroiding algorithm [11], that along with a real time knowledge of sodium profiles, would minimize the bias errors by adjusting the reference slopes in an almost real time. The Sodium profile knowledge may be inferred from the LGSWFS measurement itself [12], even thought the vertical resolution might not be exquisite because of the truncation, or from an external device if available.

- A third option could be to reduce the truncation aspect by optically shrinking the spots. Using cylindrical lenses, as introduced in Hugot et al. [13], may do this. This would probably need to be demonstrated on a bench, in order to fully validate the concept.

- Another way to reduce the truncation impact could be to work with bigger pixels (as projected on-sky) on the SH-WFS. Doing so, we are introducing non-linearities in the SH measurements, as we will be under sampled on the non-elongated direction. In that case, one can either try to follow the LGSWFS gains (subaperture per subaperture) and on an almost real time fashion (e.g. Neichel et al. [14]), or if the performance holds, simply degrade the LGS spot size as seen on sky, to match the pixel size. This can be done easily by defocusing the LGS, or having the WFS detector out-of the lenslet focus.

- An original approach, proposed by Gendron et al. [15], is to adapt the lenslet pupil sampling with the elongation pattern, and adapt the lenselet size across the pupil.

- It may always be possible to work with pyramid, even in the LGS case [16]

- Optimal tomographic reconstructor, taking explicitly into account the elongation may also be used [17,18]

- Finally, the approach that we looked at first was to simply reject the corrupted measurements, i.e. those that are strongly affected by truncation. This is simply done in the interaction matrix, before inverting it to compute the control matrix, by removing the associated lines. Once inverted, we set them back but with 0. This would certainly be a killer for single LGS systems, but with HARMONI, we have a huge redundancy in the measurements, with a large pupil overlap at all relevant altitudes.

Preliminary results for this last method are presented in Fig. 11.

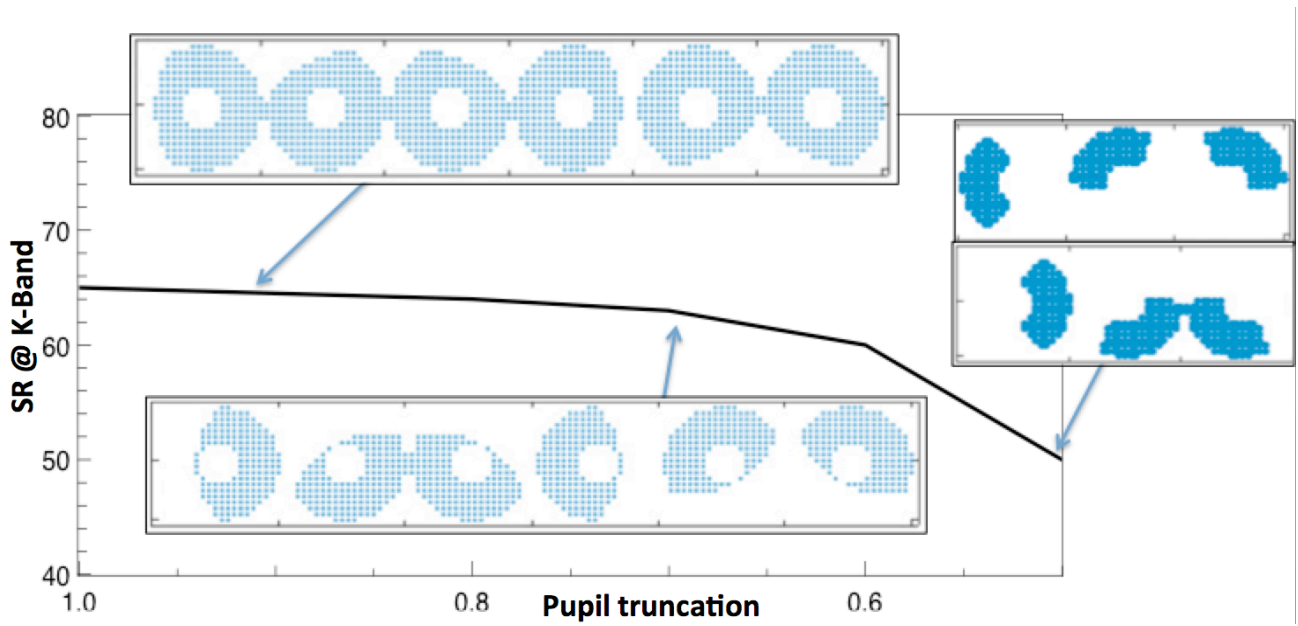

**Figure 11: LTAO performance versus pupil truncation.**

Fig. 11 shows the SR measured in K-band (only tomographic error considered here, no noise), for different level of pupil truncation. A pupil truncation of 1 means that all the measurements are kept, then the inserts show the geometrical pupil truncation that we are applying.

It is interesting to note that the performance would decrease slowly for quite some large sub-aperture rejection levels, before dropping when a large number of measurements are thrown away. Again, this is mainly possible because of the very large pupil overlap provided by the small HARMONI constellation. Those results however are to be treated as preliminary, as many error terms are not included in here, and the sensitivity of the performance to variation of Cn2, number of LGS, sodium structures etc… is to be looked at in much details. But in any cases, it provides a potential lead for an easy-to-implement working solution. And once this baseline can be demonstrated, we can always work on optimizing the performance by applying a combination of the other solutions proposed above.

### 3.4 The LTAO NGSWFS module

The LGS measurements have to be complemented with NGS measurements. This should cover Tip-Tilt at a fast frame rate (e.g. 500Hz), focus at a decent frame rate (e.g. 100Hz) and higher order modes for truth sensing at a TBD rate. The high-order update rate strongly depends on the LGSWFS design choice, hence before the LGSWFS study is fully complete, we focused on the Tipt-Tilt / Focus (TTF) performance. In the frame of HARMONI, and regarding the Top Level Requirements defined above, we have considered that working with a single star for TTF would represent a non-negligible system simplicity. Hence we have looked at the performance evolution, in terms of residual jitter, when a single NGS is used. Fig. 12 shows how the residual jitter varies when the position of this NGS moves away from the optical axis, and for different NGS magnitude. The resulting sky coverage still need to be consolidated, but the two level of performance described above would more or less corresponds to a residual jitter of respectively 2 and 5 mas (shown as red dashed line in Fig. 12).

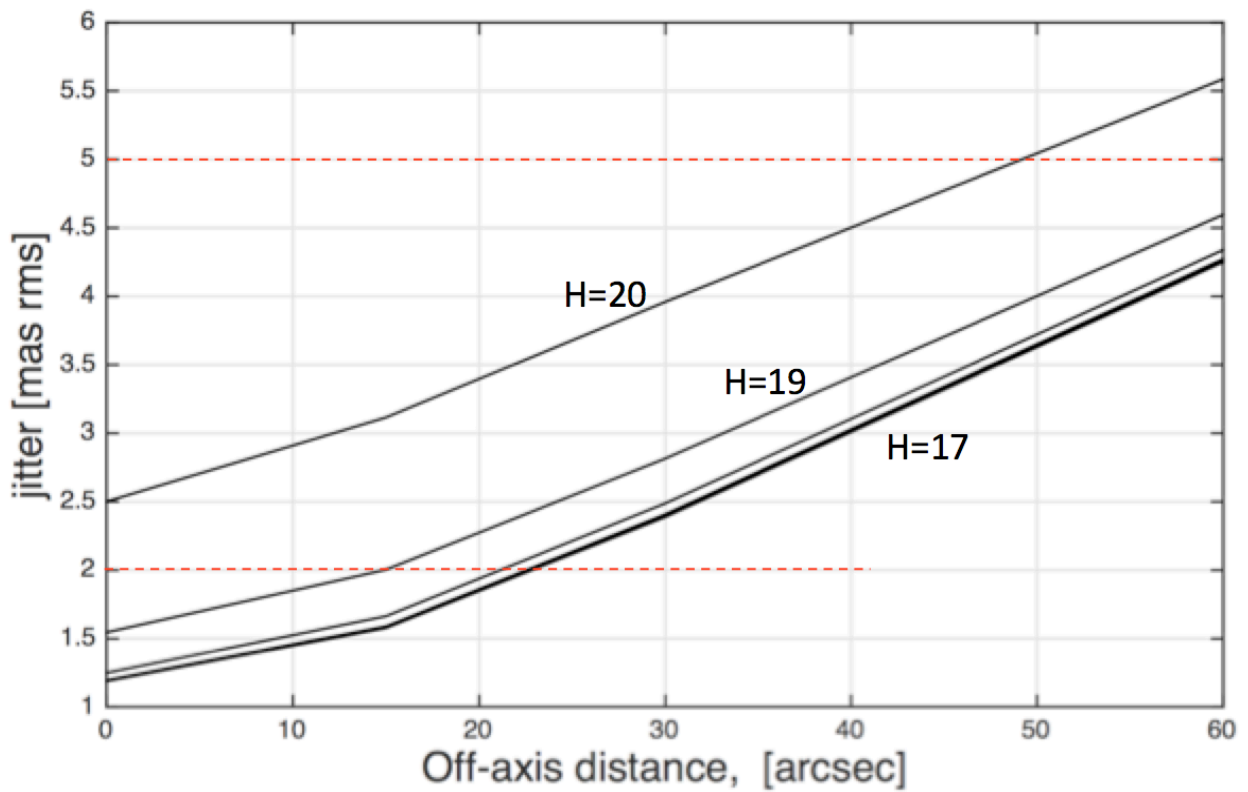

**Figure 12: residual jitter versus NGS off-axis distance for different NGS magnitudes**

The best performance (i.e. residual jitter below 2mas) would only be possible to get for close-by NGS. Note that we are considering here working with NIR NGS, to benefit from the LTAO correction. All this work is fully detailed in a partner paper: C. Correia et al. [19]

## 4. CONCLUSION

We have presented a status update of the on-going Preliminary Design Review study, carried for both an SCAO and an LTAO system to assist HARMONI observations. The current baseline for the SCAO is to work with a Pyramid WFS, and a first implementation at the interface with the HARMONI instrument is currently developed. Full E2E simulations are going-on in order to better understand the Pyramid performance, and the main limitations associated with this new type of WFS. In particular, questions regarding the robustness, the bootstrapping, the management of diffraction effects are currently addressed. On the LTAO side, the main efforts have been put on the LGSWFS aspects, as this is most likely the main challenge of the instrument. Indeed, the traditional WFS strategies will not be applicable for the E-ELT, because the available detector formats will not be adapted to the strong LGS spot elongation. Therefore, several new methods are currently tested in simulations. The impact on the tomography and the final system design are also under investigation. HARMONI should have the PDR toward the end of 2017. The instrument will then enter in a 2 years Final Design Review (FDR) phase, aiming for a first light in 2024.